\def \be {\begin{equation}}
\def \eq {\end{equation}}
\def \bee {\begin{eqnarray}}
\def \eqq {\end{eqnarray}}
\def \nn {\nonumber}
\def \bea {\begin{array}{c}}
\def \eqa {\end{array}}
\def \y {\;}

\def \ra {\rangle}

\def \Z {{\bf Z}}
\def \del {\partial}
\def \dels {\partial\kern-.5em / \kern.5em}
\def \As {{A\kern-.5em / \kern.5em}}
\def \Ds {D\kern-.7em / \kern.5em}
\def \Psib {\bar{\Psi}}

\def \a {\alpha}
\def \b {\beta}
\def \g {\gamma}
\def \d {\delta}
\def \eps {\epsilon}
\def \m {\mu}
\def \n {\nu}
\def \k {\kappa}
\def \lam {\lambda}

\def \om {\omega}

\def \th {\theta}
\def \Th {\Theta}

\def \t {\tau}

\def \hA {\hat{A}}
\def \hPsi {\hat{\Psi}}

\def \Sh {\hat{S}}
\def \Thb {\overline{\Theta}}
\def \epsb {\bar{\epsilon}}

\def \trl {\underline{tr}}

\def \II {I\hspace{-.1em}I\hspace{.1em}}
\def \IIA {\mbox{\II A\hspace{.2em}}}
\def \IIB {\mbox{\II B\hspace{.2em}}}

\documentstyle[12pt,amssymbols]{article}

\begin{document}
\begin{titlepage}
\today          \hfill 
\begin{center}
\hfill    UU-HEP/97-01\\

\vskip .5in

{\large \bf  IIB/M Duality and 
Longitudinal Membranes\\ 
in M(atrix) Theory 

}
\vskip .5in
Pei-Ming Ho and Yong-Shi Wu \\

\vskip .2in
{\em Department of Physics,
University of Utah \\
Salt Lake City, Utah 84112}
\end{center}

\vskip .8in

\begin{abstract}

In this paper we study duality properties of 
the M(atrix) theory compactified on a circle. 
We establish the equivalence of this theory 
to the strong coupling limit of 
type \IIB string theory compactified on a circle. 
In the M(atrix) theory context, our major 
evidence for this duality consists of 
identifying the BPS states of \IIB strings
in the spectrum and finding the remnant 
symmetry of $SL(2,Z)$ and the associated
$\tau$ moduli. By this \IIB/M duality, a 
number of insights are gained into the 
physics of longitudinal membranes in the 
infinite momentum frame. We also 
point out an accidental affine Lie symmetry 
in the theory. 

\end{abstract}

\end{titlepage}

\newpage
\renewcommand{\thepage}{\arabic{page}}
\setcounter{page}{1}

\section{Introduction}

The recent revelation of various string dualities 
indicates clearly that all the previously known 
five consistent perturbative superstring theories 
in ten dimensions are closely related to each 
other; they appear to represent five corners of 
the moduli space for vacuua of the one and same 
theory, which yet in another corner of the moduli 
space is most conveniently viewed as a theory
in eleven dimensions, dubbed M theory.
(For recent reviews see, e.g., 
\nocite{Sch2,Sen,Pol1,Duff,Doug1,Tow1}  
\cite{Sch2}--\cite{Tow1}.)
Banks, Fischler, Shenker and Susskind \cite{BFSS} 
have proposed a definition of M theory in the
infinite momentum frame (IMF) as a large $N$ 
limit of maximally supersymmetric quantum 
mechanics of $N\times N$ matrices describing 
D0-branes. This proposal has passed a number 
of consistency tests 
\nocite{BD,Sus,GRT,AB,LM,Doug2,BSS,Lif1,KSS,KR,Low,SS,BC,Roz} 
\cite{BFSS},\cite{BD}--\cite{Roz}.  
Many of the tests consist of verifying 
that M(atrix) theory indeed reproduces 
the right properties and interactions of 
D$p$-branes (with even $p$) expected from 
type \IIA string theory. And others examine 
the way certain string dualities are realized 
upon compactification on tori or more complicated 
spaces, and/or address some other issues.
  
The IMF description is essential to this 
formulation of M(atrix) theory, which lacks 
manifest Lorentz invariance in eleven dimensions,
and may give rise to some technical complications. 
For example, T-duality is not manifest, and 
nobody has succeeded in constructing a 
membrane wrapped in the longitudinal 
direction that defines the IMF. Our paper 
is devoted to a study of the M(atrix) theory 
compactified on a circle, and of its duality 
properties closely related to type \IIB strings,
which was motivated by an intention for 
better understanding physics in the IMF.

The basic idea is the following. The uncompactified
M(atrix) theory is supposed to be equivalent to the 
strong coupling limit of type \IIA string theory. 
Therefore, combining with the well-known \IIA/\IIB 
duality, the M(atrix) theory {\it compactified on a 
circle} should be equivalent to the {\it strong 
coupling limit} of type \IIB superstring theory
compactified on a circle. Note that the \IIB/M 
duality we are talking about here is {\it not} 
the usual one \cite{Sch1,Sch2}, which involves 
compactifying M theory on a transverse torus.
The \IIB/M 
duality examined in this paper will allow us
to infer the properties of longitudinal 
membranes. 

For this duality in the context of 
M(atrix) theory, we will provide the
following evidence. First, we will identify, 
in the IMF spectrum of M(atrix) theory, 
excitations corresponding to the BPS states of  
\IIB strings. They include oscillation
modes on the D-string resulting from compactification
which, according to 
\IIB/M duality, can be identified as 
excitations of a longitudinal membrane.
Then, we will show that relations among 
the parameters of equivalent \IIA, \IIB and 
M theories (all compactified on a circle) 
dictated by duality of each pair are 
satisfied in M(atrix) theory. 
Moreover, we will show that the remnant 
symmetry of the celebrated S-duality 
$SL(2,Z)$ for \IIB strings in the IMF is
a subgroup isomorphic to the group of 
all integers; and we identify the $\tau$ 
moduli of the strongly coupled \IIB string 
theory with the geometric tilt of the circle 
on which M(atrix) theory is compactified. 

{}Finally we present an observation that
when compactified on a circle, the M(atrix) 
theory action possesses not only a $U(N)$ 
loop group symmetry, but also the full 
affine Lie group symmetry, which is 
the central extension of the loop group, 
even at the {\it classical} level. 

{}From this study, a number of insights are 
gained into the IMF description of the 
properties of longitudinal membranes in 
M(atrix) theory. Many of them can be infered
from the properties of the D-strings.
(At the end of
the paper, an appendix is devoted to the 
description of moduli-dependent aspects of 
compactification on a slanted transverse torus.)

\section{\IIB/M Duality Revisited}
\label{IIB/M}

Let us first consider what we would expect 
to happen in M theory, in accordance with
\IIB/M duality \cite{Sch,Sch1}, when 
compactified on a transverse circle in the IMF.
 
The spectrum of the type \IIB theory 
compactified on a circle of radius $R_B$ for 
a string of NS-NS and RR charges $(q_1, q_2)$, 
with $q_1,q_2$ mutually prime, is
\be \label{MB}
M_B^2=\left(\frac{n}{R_B}\right)^2
+\left(2\pi R_B T_{(q_1,q_2)} l\right)^2
+4\pi T_{(q_1,q_2)}\left(N_R+N_L\right).
\eq
The first term is the contribution from
the Kaluza-Klein excitations, the second 
term from the winding modes and the third 
term from the string oscillation modes.
The tension of the $(q_1,q_2)$ string is
\be
T_{(q_1,q_2)}=\left((q_1-\chi_0 q_2)^2
+g_B^{-2}q_2^2\right)^{1/2} T_s,
\eq
where $\chi$ is the axion field and the subscript $0$
stands for its vacuum expectation value;
$T_s$ is the fundamental string tension in
string metric (with $l_s$ the string length
scale)
\be \label{tension}
T_s=\frac{1}{2\pi\a'}=\frac{1}{2\pi l_s^2}.
\eq
Using the level matching condition $N_R-N_L=nl$
and the BPS condition $N_R=0$ or $N_L=0$,
one finds the spectrum of \IIB BPS states 
\be \label{MB1}
M_B=\frac{n}{R_B}+2\pi R_B T_{(q_1,q_2)} l.
\eq

The spectrum of the M theory compactified
on a torus of modular parameter $\t=\t_1+i\t_2$
with radii $R_1$ and $R_{11}$ ($\t_2=R_1/R_{11}$) is 
\cite{Sch1,Sch2}
\be \label{MM}
M_M^2=\left(\frac{N}{R_{11}}\right)^2
+\left(\frac{m-\t_1 N}{R_1}\right)^2
+\left(A_M T_2^M n\right)^2+\cdots,
\eq
where $A_M=(2\pi R_{11})(2\pi R_1)$
is the area of the torus.
The first two terms come from the Kaluza-Klein 
modes, the third from the winding modes and
the contribution of membrane excitations
are not written down because the quantum theory
of the membrane is what we are going after.
Comparing the spectra (\ref{MB}) and (\ref{MM}), 
one finds that the Kaluza-Klein (winding)
modes in type \IIB theory match with
the winding (Kaluza-Klein) modes in M theory,
if we make the identification $m=q_1 l$, 
$N=q_2 l$, and
\footnote{Throughout this paper all quantities 
appearing in the same equation
are given with respect to the same metric.}
\bee
&R_B=1/(A_M T_2^M), \label{RBAT}\\
&R_B=l_s^2/R_1, \label{RB}\\
&\chi_0=\t_1, \label{t1}\\
&g_B=R_{11}/R_1(=1/\tau_2). \label{RRg}
\eqq
Therefore the modular parameter of the torus
is identified with the vacuum expectation value
of the complex field $\chi+ie^{-\phi}$,
where $\phi$ is the dilaton field and $g_B=e^{\phi_0}$.

Here the consistency of eqs. (\ref{RBAT}) and 
(\ref{RB}) requires the relation 
\be\label{RBT}
T_s=2\pi R_{11}T_2^M,
\eq
implying that a fundamental \IIB string is identified 
with a membrane wrapped on $R_{11}$. Therefore, 
given the parameters of M theory, the paramenters 
of type \IIB theory are determined by eqs. 
(\ref{RBAT}), (\ref{t1}-\ref{RBT}). Conversely, 
given the parameters of type \IIB theory, the 
parameters of M theory are determined by:
\be
R_1=\frac{l_s^2}{R_B}, \quad
R_{11}=g_B\frac{l_s^2}{R_B}, \quad
T_2^M=g_B^{-1}R_B T_s^2. \label{TRBT}
\eq

This duality also matches a membrane wrapped 
around a cycle of the torus in M theory with a 
string in type \IIB theory: A cycle on the torus 
is specified by two mutaully prime integers 
$(q_1,q_2)$ with a minimal length
\be \label{Lqq}
L_{(q_1,q_2)}=2\pi R_{11}
\left((q_1-\tau_1 q_2)^2+\tau_2^2 q_2^2
\right)^{1/2}.
\eq
Hence the tension of a \IIB string
of charge $(q_1,q_2)$ is
$T_{(q_1,q_2)}=L_{(q_1.q_2)}T_2^M$,
in agreement with the above relations.

Once the \IIB/M duality is justified, the 
spectrum for M theory can be completed
by looking at the \IIB spectrum (\ref{MB1}) 
\cite{Sch2}:
\be
M_M=\left(\left(\frac{N}{R_{11}}\right)^2
+\left(\frac{m-\t_1 N}{R_1}\right)^2\right)^{1/2}
+A_M T_2^M n \, ,
\eq
for arbitrary integers $N$, $m$ and $n$.
Comparing this with eq. (\ref{MM}), we see that
the second term includes both the contributions 
from winding modes and excitations on the membrane.
To compare this spectrum to that of M(atrix) theory,
we need to go to the IMF by boosting in 
the direction of $R_{11}$ so that $P_{11}\equiv
N/R_{11}\gg m/R_1$. (This Lorentz transformation 
corresponds to a change in the RR-charge $q_2$ 
in type \IIB theory to a large number, which  
can be achieved by an $SL(2,\Z)$ symmetry 
transformation.) 
Thus, in the IMF, the spectrum in M theory
appears to be
\be \label{MM1}
M_M=\frac{N}{R'_{11}}
+\frac{R'_{11}}{2N}\left(\frac{m}{R'_1}\right)^2
-\frac{\t_1 m}{\t_2 R'_1}+A_M T_2^M n+\cdots,
\eq
where (see Fig.1)
\be \label{R1'}
R'_{11}=\frac{\t_2}{|\t|}R_{11}, \quad
R'_1=\frac{|\t|}{\t_2}R_1.
\eq

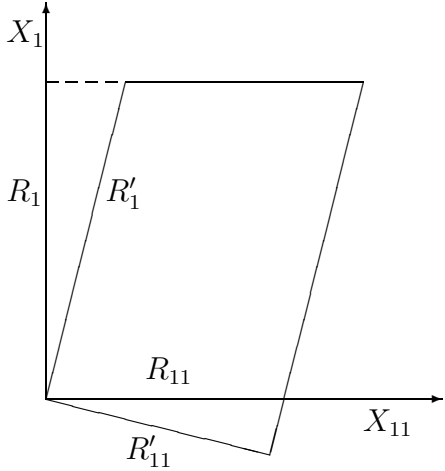
\begin{figure}[h]
\setlength{\unitlength}{1.5pt}
\hspace{1.5in}
\vspace{0in}
\begin{picture}(100,120)(-20,-20)
\put(0,0){\vector(0,1){100}}
\put(0,0){\vector(1,0){100}}
\put(0,0){\line(1,4){20}}
\put(20,80){\line(1,0){60}}
\put(80,80){\line(-1,-4){23.53}}
\put(0,0){\line(4,-1){56.47}}
\multiput(0,80)(5,0){4}{\line(1,0){3}}
\put(-10,90){$X_1$}
\put(80,-8){$X_{11}$}
\put(-10,50){$R_1$}
\put(15,50){$R'_1$}
\put(25,5){$R_{11}$}
\put(20,-15){$R'_{11}$}
\end{picture}
\vspace{0in}
\caption{Longitudinal Torus}
\vspace{0in}
\end{figure}


We note that the fourth term in eq. (\ref{MM1})
is finite in the limit $N\rightarrow\infty$,
without a factor of $1/P_{11}$ in front of it,
showing that
the energy of excitations on a longitudinal 
membrane scales as $P_{11}$ under a boost in 
the longitudinal direction \cite{BSS}.

In next section we will try to identify the
 spectrum (\ref{MM1}) in the M(atrix) 
theory compactified on a circle. 

\section{M(atrix) Theory and Compactification}
\label{matrix}

In this section, we first review the uncompactified
M(atrix) theory, with an eye on the guiding role
of  \IIA/M duality and on its IMF nature. Then, 
we will describe compactification of  the theory 
on a circle, {\it not necessarily perpendicular} to 
the longitudinal direction that defines the IMF. 

\subsection{\IIA/M Duality and Uncompactified 
M(atrix) Theory}
\label{uncomp}

The BFSS action for M theory in the IMF 
is a large $N$ limit of the supersymmetric 
matrix quantum mechanics obtained by 
dimensionally reducing the supersymmetric 
$U(N)$ Yang-Mills action from $9+1$ 
dimensions to $0+1$ dimension \cite{BFSS}:
\bee \label{D0}
S&=&\frac{1}{g}\int dt\;\;
tr\left(-\frac{1}{2}\nabla_0 A_i\nabla_0 A_i
+\frac{1}{4}[A_i,A_j][A_i,A_j]\right. \nn \\
&&\left.+\frac{i}{2}\Psib(\g^0\nabla_0\Psi
+\g^i[A_i,\Psi])\right),
\eqq
where $i,j=1,2,\cdots,9$ and 
$\nabla_0=\del_t+[A_0,\cdot\;]$.
$A_{\m}$ and $\Psi$ are anti-hermitian 
$N\times N$ matrices; $A_{\m}^a$ are 
real and $\Psi^a$ are Majorana-Weyl 
spinors in 10 dimensions. 

This matrix action was originally 
suggested \cite{dWNH} for a regularized 
supermembrane in eleven dimensions,
since the $U(N)$ gauge symmetry
\bee
A_0\rightarrow UA_0 U^{\dagger}
+U\del_t U^{\dagger}, \quad
A_i\rightarrow UA_i U^{\dagger}, \quad
\Psi\rightarrow U\Psi U^{\dagger},
\eqq
can be interpreted as the area-preserving
diffeomorphism group of a membrane
in the large $N$ limit.

In the context of string theory, this action 
describes dynamics of $N$ D0-branes in type 
\IIA theory \cite{Wit}. In the temporal gauge 
$A_0=0$, using the hermitian matrices $X_i=
-iA_i$ $(i=1,2,\cdots 9)$, the Hamiltonian is
\be
H=tr\left(\frac{g}{2}\Pi_i^2-\frac{1}{4g}[X_i, X_j]^2
+\frac{1}{2g}\Psib\g^i[X_i, \Psi]\right).
\label{Hamil}
\eq
A minimum of the potential term in the Hamiltonian
is reached when $\Psi=0$ and all the $X_i$'s can
be simultaneously diagonalized. Introducing
\be \label{Xx}
X_i=T_s x_i, \quad\quad\quad
\eq
then $(x_i)_{\a\a}$ ($\a=1,2,
\cdots N$) is interpreted as the $i$-th 
coordinate of the $\a$-th D0-brane.
An off-diagonal entry $(x_i)_{\a\b}$ ($\a\neq\b$)
represents the effects due to open strings stretched
between the $\a$-th and the $\b$-th D0-brane.
Hence the energy of a stretched 
string, given by the string tension times the 
distance between the D0-branes, should equal the 
mass of the field $(x_i)_{\a\b}$ in the action 
\cite{Wit}. The coefficient of the action (\ref{D0})
is thus $1/g=T_0/T_s^2$, where $T_0$ is 
the D0-brane tension (the D-particle mass).

In accordance with the well-known \IIA/M duality 
between the M theory compactified on $S^1$ 
(with radius $R_{11}$) and type \IIA theory 
\cite{Tow,Wit2,Sch1}, membranes wrapped around 
$S^1$ are identified with fundamental \IIA strings, 
and unwrapped membranes with D2-branes. 
The string tension and D2-brane tension are
therefore related to the membrane tension $T_2^M$ 
by $T_s=2\pi R_{11}T_2^M$ and $T_2=T_2^M$.
Recall that the D$p$-brane tension in  type \II 
string theory is given by \cite{PCJ,dAl}
\be
T_p=1/((2\pi)^p g_s l_s^{(p+1)}), 
\eq
where $g_s$ is the string coupling $g_A$ ($g_B$)
in type \IIA (\IIB) theory for $p$ even (odd).
Therefore the compactification radius $R_{11}$ and the 
membrane tension $T_2^M$ can be given in terms
of the \IIA parameters $(g_A, l_s)$ as
\be \label{MIIA}
R_{11}=g_A l_s, \quad T_2^M=
\frac{1}{(2\pi)^2 g_A l_s^3},
\eq
or conversely,
\be \label{MIIA1}
g_A=2\pi R_{11}^{3/2}(T_2^M)^{1/2}, \quad
l_s=\frac{1}{2\pi R_{11}^{1/2}(T_2^M)^{1/2}}.
\eq
(The Planck length $l_p$ in M theory is
defined by $T_2^M=1/((2\pi)^2 l_p^3)$,
implying that $l_p=g_A^{1/3}l_s$.)
It follows that
\be \label{gR}
\frac{T_s^2}{g}=T_0= \frac{1}{R_{11}}.
\eq

According to eq. (\ref{Hamil}), the 
center-of-mass kinetic energy 
of  $N$ D0-branes is 
\be \label{P2}
\frac{R_{11}}{2N}\sum_i(p_i^{com})^2,
\eq
where $p_i^{com}$ is the conjugate 
momentum of the center-of-mass position 
$x_i^{com}=tr(x_i)/N$. The prefactor
has an interpretation in M theory
as the momentum of $N$ partons (D0-branes) in 
the compactified direction: $P_{11}=
N/ R_{11}$.  So the Hamiltonian
(\ref{Hamil}) is understood as
the light-cone energy with $P_{11}$ very
large, implying the eleven dimensional
Lorentz invariance with the IMF 
expansion of kinetic energy:
$K=(P_{11}^2+\sum P_i^2)^{1/2}
=P_{11}+(\sum_i P_i^2)/2P_{11}+\cdots$.

In the limit $R_{11} \to \infty$ with $l_p$ 
kept fixed, type \IIA theory goes to the strong 
coupling limit while its dual theory -- M theory --
becomes uncompactified in eleven dimensions and 
are dominated by massless D0-branes (supergravitons).
{\it Guided by the \IIA/M duality}, the BFSS 
formulation of M(atrix) theory \cite{BFSS} just  
postulates that in the limit both $R_{11}$ and 
$N/R_{11}$ going to infinity, the Hamiltonian 
(\ref{Hamil}) describes the uncompactified 
M theory in the IMF.
 
Here $R_{11}$ is used as an infrared 
cutoff for the uncompactified theory: 
All winding modes are supposed to be 
thrown away, except those which wind 
$R_{11}$ at most once corresponding
to longitudinal branes. 

To test the \IIB/M duality we revisited  in 
last section, normally one would consider 
compactification of M(atrix) theory on a 
transverse torus \cite{BFSS,Tay,Sus,GRT,SS},
which needs to examine a $(2+1)$-dimensional 
quantum gauge theory. However, as we 
are going to demonstrate, 
it is more instructive to test \IIB/M 
duality in the M(atrix) theory compactified 
only on a {\it circle} which,
by combining the usual \IIA/\IIB and \IIA/M dualities,
is expected to be the strong coupling
limit of type \IIB string theory. One expects 
to gain interesting insights into physics in
the IMF formulation from this study, because 
it will involve longitudinal membranes 
that wrap around $R_{11}$.

\subsection{M(atrix) Theory Compactified 
on an Oblique Circle}
\label{comp}

Now let us consider the M(atrix) theory 
compactified on a circle, which is normally 
\cite{BFSS,Tay,GRT} taken to be in a 
direction, say $X_1$, perpendicular to the 
longitudinal $X_{11}$, with radius $R_1$. 
For our purpose, it is necessary to incorporate
moduli parameters for equivalent \IIB string 
theory, which needs to consider the 
compactification on an {\it oblique} circle, 
in a tilted $X'_1$-direction in the
$X_1-X_{11}$ plane, with radius $R'_1$.
This makes sense in M(atrix) theory. From
the target-space point of view, at least in
low energy supergravity, what is relevant is 
the Kaluza-Klein metric \cite{Sch1,Sch2} which 
gives rise to the moduli parameters of \IIB 
strings. From the world-volume point
of view, the resulting D1-brane action will be
defined on the dual circle, whose
radius is essential for quantization of  the
momentum modes of the D1-brane.

First recall the usual case with $S^1$ in the
$X_1$-direction. By gauging a discrete 
subgroup of $U(N)$ representing periodic 
translations in $X_1$, the D0-brane action in
the compactified space can be written as 
a D1-brane action on the dual circle 
\cite{BFSS,Tay,GRT}. In accordance with 
\IIA/\IIB T-duality, the winding modes 
around the circle for open strings stretched 
between D0-branes become the discretized 
momentum modes for D1-branes on the dual 
circle, while the compactified coordinate $X^1$ 
of the target space turns into
the covariant derivative with a gauge field 
$A_1$ on the worldsheet, leading to the action 
(in the temporal gauge) \cite{Tay}:
\bee \label{D1}
S&=&\int dt \int_0^{2\pi R_B}\frac{dx}{2\pi R_B}
\;\frac{1}{2g}\;
tr\left(\dot{X}_i^2-\dot{A_1}^2-(\nabla_1 X_i)^2
+\frac{1}{2}[X_i, X_j]^2\right. \nn\\
&&\left.+i\Psib(\g^0\dot{\Psi}
+\g^1\nabla_1\Psi+\g^i[iX_i, \Psi])\right),
\eqq
where $i=2,3,\cdots,9$,
$\nabla_1=\del_x+[A_1,\cdot\;]$ and
$R_B=l_s^2/R_1$ is the dual radius of $R_1$,
exactly as required by \IIA/\IIB duality. 

Now let us consider the case of an oblique 
$S^1$ in a tilted $X'_1$-direction.
In low-energy supergravity, the Kaluza-Klein
metric in such coordinates will contain a 
gauge field in the $X'_1$-direction. This 
gauge field can be gauged away except its 
Wilson line (or holonomy) degree of freedom. 
Motivated by this observation, in the dual
description, we expect the appearance of 
a constant background electric field $E$
for the gauge field on the D1-brane 
world-sheet so that 
the field strength 
$F_{01}$ is shifted to $(F'_{01}-E)$, which 
is $(\dot{A'_1}-E)$ in the temporal gauge. 
Now we denote the gauge field as $A'_1$, 
because it corresponds to the tilted 
$X'_1$-coordinate and has a period $R'_1$. 
Indeed we can see how this modification 
comes about by considering compactification 
on a slanted torus.
This leads to (see Appendix) the 
following modified action:
\bee \label{D1_1}
S&=&\int dt \int_0^{2\pi R_B}\frac{dx}
{2\pi R_B}\;\frac{1}{2g'}\; tr\left(
\dot{X}_i^2-(\dot{A'_1}-E)^2-(\nabla_1 X_i)^2
+\frac{1}{2}[X_i, X_j]^2\right. \nn\\
&&\left.+i\Psib(\g^0\dot{\Psi}
+\g^1\nabla_1\Psi+\g^i[iX_i, \Psi])\right),
\eqq
where $\nabla_1=\del_x+[A'_1,\cdot\;]$;
$R_B=l_s^2/R_1$ is unchanged, while
\bee \label{rel1'}
E= -i \lambda T_s \equiv -i
\frac{\tau_1}{\tau_2} T_s, \quad
\frac{T_s^2}{g'}=\frac{1}{R'_{11}}. 
\eqq
Here $\tau=\tau_1+i\tau_2$ is the modular 
parameter of the slanted longitudinal torus 
with radii $R_1$ and $R_{11}$; the relations 
between ($R'_1$, $R'_{11}$) and ($R_1$, 
$R_{11}$) are exactly those of eq. (\ref{R1'}).
Note that the coefficient $g'$ is defined
as (\ref{gR}) with $R_{11}$ replaced by
$R'_{11}$, while the change of $(R_1, R_{11})$ 
to $(R'_1, R'_{11})$ does not affect the area 
$A_M$. Moreover, here the values of the field 
$A'_1$ are ranged between $0$ and 
$2\pi R'_1 T_s^2$, with $R'_1$ just the radius 
of the $(0,1)$ cycle on the slanted torus 
(see eq. (\ref{Lqq})). Using eqs. (\ref{RB}) and 
(\ref{Xx}), one finds that if eq. (\ref{RRg}) 
is used to define $g_B$, then the prefactor 
of the action  (\ref{D1_1}) is precisely
the tension $T_{(0,1)}=|\t|T_s$, appropriate
for a \IIB string of charge $(0, 1)$.

Comparing this to the action (\ref{D1}), we see 
that the term $\dot{A}_1^2$ has been replaced by 
$(\dot{A'_1}-E)^2$, with the background gauge 
field $E$ given by the above relation. In 
addition to a constant term, this modification
leads to adding a topological term of the
form $-i \lambda \int \dot{A'_1}$ to the 
D1-brane action. This is an analogue of the
$\theta$-term in two dimensional gauge theory
\cite{Wit3}. 

By imposing periodic boundary 
conditions in time, a change in $\lambda$ by 
$b$ results in a change in the action by 
$-i 2\pi|\t|^2 b/\t_2$ times an integer. So 
the period for $\lambda$ in the path integral 
measure $e^{iS}$ is $\t_2 /|\t|^2$.

We note that this moduli-dependent, 
topological term recently also appears in
ref. \cite{APPS} for the action of a 
D1-brane. It was used there to 
recover the fundamental string action 
by electric-magnetic duality. Here in
our treatment this term is just right, see 
next section, to reproduce correctly 
the moduli-dependent term, the third term
in eq. (\ref{MM1}), in the IMF spectrum
of M theory. We also note that the action 
(\ref{D1_1}) has an additional constant 
term proportional to $E^2$, which is just 
right for the Hamiltonian to have a minimum 
of zero energy, in consistency with 
unbroken supersymmetry.

\section{\IIB/M Duality in M(atrix) Theory}
\label{duality}

We propose to interpret the M(atrix) theory 
action (\ref{D1_1}) as describing a system 
of $N$ D-strings in (strongly coupled) type 
\IIB theory compactified on a circle. 

\subsection{Spectrum of \IIB BPS States}
\label{spectrum}

As evidence for this equivalence,
we now show that the spectrum 
(\ref{MM1}) expected from \IIB/M 
duality can be found in the M(atrix) theory.

First, the kinetic energy of 
the $U(1)$ factor of $A'_1$, $tr(A'_1)/N$, 
gives the second term in (\ref{MM1}),
which is simply part of the matching 
(\ref{P2}) mentioned in last section.
In addition, the contribution of the 
topological term (due to the Wilson 
line) to the Hamiltonian just matches 
the third term in (\ref{MM1}) as well, 
because of the relation (\ref{rel1'}) 
between $E$ and $\t_1/\t_2$.
 

Now consider the configurations which
satisfy $[X_i,X_j]=0$ and $\Psi=0$ where
one can simultaneously diagonalize the 
$X_i$'s. The action (\ref{D1_1}) becomes 
proportional to the free action 
$\sum_{\a}\left((\dot{X}_i)_{\a\a}^2
-(\del_x X_i)_{\a\a}^2\right)$.
By Fourier expansion:
$X_i=\sum_k X_{ik}e^{ikx/R_B}$,
one finds that the excitation 
(or oscillation) modes of $X_i$ 
on the closed D1-brane (D-string) 
have the spectrum $M=n/R_B=2\pi T_s R_1 n$ 
upon quantization. The operator $n$ is 
defined by $n=\sum_{ik\a} kN_{ik\a}$,
where $N_{ik\a}$ is the number operator
for the mode $(X_{ik})_{\a\a}$. When 
$[X_i,X_j]^2$ and other terms are 
included, there are interactions 
between D-strings. However, we expect 
a non-renormalization theorem for 
the spectrum of these states, because
they correspond to BPS states of \IIB 
strings. Obviously this part of energy 
should be identified with the fourth 
term in (\ref{MM1}), or equivalently 
with the first term in (\ref{MB1}).

We observe that in the present IMF formulation, 
this part of the energy, which involves the 
excitations of the D-string, does {\it  not} 
have a prefactor of $1/(2P_{11})$, in contrast 
to the usual IMF energies for purely transverse
excitations. This
implies that the proper interpretation of this
part of energy in M theory should be attributed to
excitations on a (longitudinal) membrane 
wrapped on the eleventh direction, just as
expected from \IIB/M duality. Note 
that in this argument we have taken advantage 
of the IMF formulation, without the need of 
constructing a semi-classical longitudinal membrane.    

One can check that the topological configuration of $A'_1$
and the oscillation modes of $X_i$ are both indeed BPS states
in M(atrix) theory.
\footnote{
We thank M. Li for discussions on this matter.}
The eleven dimensional supersymmetry transformations in the IMF
consist of the dynamical part
\be \label{SU1}
\d A_{\m}=\frac{i}{2}\bar{\eps}\g_{\m}\Psi, \quad
\d\Psi=-\frac{1}{4}F_{\m\n}\g^{\m\n}\eps,
\eq
where $F^{0i}=\dot{X^i}$, $F^{1i}=\nabla_1 X^i$ and $F^{ij}=[X^i,X^j]$,
and the kinematical part
\be \label{SU2}
\d A_{\m}=0, \quad \d\Psi=i\tilde{\eps},
\eq
where $\eps$ and $\tilde{\eps}$ are both Majorana-Weyl spinors
times a unit matrix.
Each part of the supersymmetry has 16 generators
and the total supersymmetry has 32 generators.
It turns out that the supersymmetry algebra involves central terms
which can be interpreted as various RR charges \cite{BSS}.
The topological configuration of $A'_1$ preserves one half of the
total SUSY as a linear combination of
the dynamical part and the kinematical part.
The associated charge is simply the Kaluza-Klein
momentum $P_1$ proportional to $m$.
The purely left-moving (or right-moving) oscillation modes
preserve one quarter of the total SUSY
(half of the dynamical part) and give
one nonzero RR charge $Z^1$ (defined in Ref.\cite{BSS}
as a central term in the SUSY algebra)
proportional to $k$,
corresponding to
branes wrapped around $R_{11}$.
 
It is instructive to compare the above 
identification of the spectrum for a longitudinal
membrane with that for a transverse membrane 
\cite{BFSS}. The configuration in M(atrix) 
theory that represents a membrane wrapped 
$n$ times around a transverse torus with radii 
$R_1$ and $R_2$ is given by $x_1=R_1 p$,
$x_2=R_2 q$, with $p$ and $q$ satisfying
$[p,q]=2\pi i n/N$.
(While the representation of the canonical 
commutation relation can only be realized 
in infinite dimensional representations,
the commutation relation above makes sense
with the understanding that $x_i$ ($i=1,2$)
lives on a circle of radius $R_i$ and that 
the right hand side is appropriately 
normalized by the dimension of the 
representation $N$. \footnote{
We thank B. Zumino for pointing this out to us.
}) 
As easy to verify, it is the potential term 
$\frac{1}{4}[x_i,x_j]^2$ in the light-cone 
Hamiltonian (\ref{Hamil}) that correctly 
reproduces the membrane spectrum  
$A_M T_2^M n$,  where $A_M=(2\pi R_1)
(2\pi R_2)$ is the area of the torus, for this
configuration.
 
In this discussion, to match the \IIB spectrum,
it is necessary to have the membrane wound
around the $R_{11}$ only once. This is
in agreement with the use of $R_{11}$ 
in M(atrix) theory as a cutoff, so that longitudinal 
branes are allowed to wind it only once!
In addition, this is exactly what has
been conjectured by Schwarz 
in his discussions on \IIB/M duality \cite{Sch2},
namely that in M theory the membrane wrapped on a 
torus should select a preferred cycle in which 
it is wrapped many times, and this preferred 
direction must be the one defined by the \IIB theory 
Kaluza-Klein excitations, which is nothing but 
the $R_1$ direction!
This can be argued as follows.
The \IIB Kaluza-Klein modes by T-duality
are \IIA string winding modes around $R_1$.
The \IIA strings are membranes wound around $R_{11}$
only once by \IIA/M duality.
Hence the \IIB Kaluza-Klein modes are
membranes wound around $R_{11}$ once
and $R_1$ an arbitrary number of times.

\subsection{Relations among \IIA/\IIB/M Parameters}
\label{parameter}

The above spectrum matching can also be 
understood from \IIA/\IIB and \IIA/M dualities. 
Let us recall that the quantity $(X_{ik})_{\a\a}$ 
has a dual interpretation in \IIA or \IIB language.
In \IIA language, when compactifying D$0$-branes 
on a circle of radius $R_1$, an open string 
wound $k$ times around the circle with both 
ends on the same D$0$-brane labelled $\a$ 
has the energy $2\pi k T_s R_1$. Such a winding 
mode of an open string is known to be represented 
by $(X_{ik})_{\a\a}$ \cite{Tay}. On the other 
hand, in the dual (\IIB) language, $(X_{ik})_{\a\a}$ 
represents an oscillation mode
on the D-string, with energy $k/R_{B}$ which
is identical to  $2\pi k T_s R_1$.
On the other hand, 
because strings in type \IIA theory are 
thought of, by \IIA/M duality, as membranes 
wound around $R_{11}$, so we are again
led to the identification in the last subsection 
of the D-string oscillation modes with
excitations on the longitudinal membrane.

Thus, what we have here is \IIA/\IIB/M triality,
i.e. the M(atrix) theory compactified on a circle
is equivalent to either the strong coupling limit
of type \IIA or type \IIB theory, each compactified 
on a circle too. The \IIA/M duality and \IIB/M 
duality we considered separately in the above are 
linked by the usual \IIA/\IIB duality. 

If we consider the \IIA/M duality for the M theory
compactified on a torus and \IIA on a circle,
the \IIA/M relations for this case are simply
(\ref{MIIA}-\ref{MIIA1})
together with
\be
R_A=R_1,
\eq
where $R_A$ is the radius of the circle in \IIA theory.
The \IIA/\IIB duality identifies D$p$-branes in \IIA
with D$(p\pm 1)$-branes in \IIB theory
by wrapping or unwrapping.
Therefore we have (\ref{RB}) and
\be
g_B=g_A\frac{l_s}{R_A}.
\eq
It can be easily checked that these 
relations, required by dualities of 
each pair of \IIA, \IIB and M theories,
are satisfied in the M(atrix)
theory compactified on a circle. Also
using any two of the dualities,
one can derive uniquely the third one.
If one uses the dualities to go cyclicly 
from one theory through the other two 
to return to the original theory,
one finds that the parameters are 
unchanged after the journey. If it 
were not the case it would mean that 
there exist new self-dualities in 
these theories.

Because $R_{11}$ is to be treated as a cutoff,
it should be much larger than any other length scale
in the theory. Hence in the 
limit $R_{11}\rightarrow\infty$,
the matrix model gives the $S^1$-compactified
M theory dual to the type \IIB theory
compactified on the dual circle
in the strong coupling limit
according to (\ref{RRg}).

\subsection{$SL(2,\Z)$ Duality of \IIB Theory}
\label{SL2Z}

The $SL(2,\Z)$ duality of \IIB theory
transforms a $(q_1, q_2)$ string to
a string with charge $(q'_1, q'_2)$ given by
\be \label{qqmN}
\left(\begin{array}{c}
q_1 \\
q_2 \end{array}\right)\rightarrow
\left(\begin{array}{c}
q'_1 \\
q'_2 \end{array}\right)=
\left(\begin{array}{cc}
   a & b \\
   c & d
\end{array}\right)
\left(\begin{array}{c}
q_1 \\
q_2 \end{array}\right),
\eq
where $\left(\begin{array}{cc}
   a & b \\
   c & d
\end{array}\right)$
is an $SL(2,\Z)$ matrix.
The coupling and string tension transform as
\bee
&g_B\rightarrow g'_B=|c\t+d|^2 g_B, \\
&T_s\rightarrow T'_s=|c\t+d| T_s.
\eqq

These agree, by the \IIB/M duality 
relations (\ref{RBAT}-\ref{TRBT}),
with the modular transformation
\be \label{modular}
\t\rightarrow \t'=\frac{a\t+b}{c\t+d}
\eq
of the torus in M theory, on which the
theory is compactified. For the 
modular transformation to be a 
geometric symmetry of M theory,
the area $A_M$ of the torus is to be fixed
and thus the radii transform as
\be \label{R1R11}
R_1\rightarrow \frac{R_1}{|c\t+d|}, \quad
R_{11}\rightarrow |c\t+d| R_{11}.
\eq
The spectrum in M theory,
$\left((N/R_{11})^2+((m-\t_1 N)/R_1)^2\right)^{1/2}$
is invariant under the transformation
(\ref{modular}), (\ref{R1R11}) and
\be \label{qqmN1}
\left(\begin{array}{c}
m \\
N \end{array}\right)\rightarrow
\left(\begin{array}{c}
m' \\
N' \end{array}\right)=
\left(\begin{array}{cc}
   a & b \\
   c & d
\end{array}\right)
\left(\begin{array}{c}
m \\
N
\end{array}\right).
\eq

In the IMF description of the M(atrix) theory
compactified on a circle, we do not expect to
have the full $SL(2,Z)$ symmetry, since
the longitudinal direction is preferred, and
the limit of $R_{11}$ may be different from that
of $R_1$. However, the theory may be invariant
under a subgroup. We have checked that if
we make the transformation (\ref{modular}-\ref{qqmN1}),
then the invariance of the IMF spectrum 
(\ref{MM1}) requires that $a=1,b=0$ and
$d=1$. So the remnant symmetry is
\be
\left(\begin{array}{cc}
         1 & 0 \\
         c & 1
      \end{array}\right)
\eq
where $c$ is an integer.

\subsection{Finiteness of Spectrum and Limits of Parameters}

For the spectrum $M_M$, eq. (\ref{MM1}), we obtained 
in Sec.\ref{spectrum} for M(atrix) theory 
to make sense in the limit 
$R_{11}\rightarrow\infty$,
other parameters in the theory have to
take appropriate limits accordingly.
To be more precise about these limits,
one should consider only dimensionless quantities.
For instance,
\be
r_B=R_B/l_s, \quad 
r_B^c=g_B^{-1/4}R_B/l_s
\eq
are the \IIB radius measured in
the \IIB string metric and the canonical metric,
where $l_s=1$ and $g_B^{1/4}l_s=1$, respectively;
\be
r_1=R_1/l_p, \quad
r_{11}=R_{11}/l_p
\eq
are
the values of $R_1$ and $R_{11}$
measured in the eleven-dimensional Planck units.
Similarly, the finiteness of the spectrum
is to be considered with respect to a certain 
system of units.

In the M theory it is natural to measure 
everything in terms of the Planck scale 
$l_P$, so the dimensionless spectrum is 
$m_P=M_M l_P$. In \IIB theory one can 
choose to use the string metric or the 
canonical metric, where the spectrum 
appears to be $m_s=M_M l_s$ and 
$m_s^c=g_B^{1/4}M_M l_s$, respectively.
The results in the \IIB canonical metric 
are the same as in the Planck units.
In the following we discuss separately
in the Planck units and string units
the appropriate limits of
various parameters in the theory
for the spectrum to be finite.

In the Planck units,
for both the third and the fourth 
terms in eq. (\ref{MM1}) to have finite 
limits, we need 
\be
r_B^c\sim\mbox{finite}, \quad r_1\sim r_{11}^{-1}, 
\quad \t_1\sim r_{11}^{-3}, \quad
\t_2\sim r_{11}^{-2}.
\eq
As a consequence, the parameter $\lambda\equiv
\t_1/\t_2$ does not have a
finite period since its period $\t_2/|\t|^2$
goes to infinity.
It is also easy to
check that the modular parameter 
$\t_1/\t_2$ of our longitudinal torus,
scaled by $(l_p/l_s)^2$ so that it
has a finite limit, is
invariant under the remnant subgroup of $SL(2,\Z)$
mentioned in the previous subsection.
In this case one has $r_1\rightarrow 0$.
This gives a version of the M theory 
dual to the \IIB theory compactified 
on a finite circle in the strong coupling limit.
The part of the spectrum 
considered above has a finite limit in both 
the Planck units and the \IIB canonical metric.
The opposite extremum $r_1\rightarrow\infty$
is just the original case considered in 
Ref.\cite{BFSS}.

In the string metric we need
\be
r_B\sim\mbox{finite}, \quad \t_2\sim r_{11}^{-3/2}
\eq
for arbitrary $\t_1$.
In this case it is possible to choose $\t_1\sim\t_2^{1/2}$
so that the parameter $\lam$ has a finite period.
The modular parameter $\t_1/\t_2$ can have arbitrary limit
by choosing the limit of $\t_1$.
But in any case $\t_1/\t_2$ scaled by an appropriate
power of $l_p/l_s$ is invariant under the remnant subgroup.
Here we get a strong coupling limit of \IIB theory
compactified on a circle with a finite radius in string units
but zero radius in the canonical metric.

It is certainly possible to take other limits.
For instance, one can take the units such that
$R_1$ is finite, then one needs $T_2^M\rightarrow 0$. 
This is a special limit in the \IIB/M duality
which is the strong coupling limit of \IIB theory
(or by T-duality, \IIA theory) compactified on a circle
as well as the weak tension limit in the M theory
compactified on the dual circle.

\section{Affine Lie Group Symmetry}\label{ALGS}

An accidental affine Lie algebraic 
structure appears in M(atrix) theory 
when it is compactified on a circle.
In Ref.\cite{CH}, WZW models of Lie 
group valued fields on two dimensional 
spacetime are rewritten as WZW models 
of affine Lie group valued fields on one 
dimensional world history. In the same 
spirit, by imposing periodic boundary 
conditions,
\footnote{One can also impose twisted
boundary conditions which would lead to twisted
affine Lie groups.}
we decompose the $x$ 
dependence of fields into 
their Fourier modes:
\be
A_{\m}=\a_{\m m}^{\y a}T^a e^{imx/R},
\quad \Psi=\psi^a_m T^a e^{imx/R},
\eq
where $R=R_B$.
Since $A_{\m}$ and $T^a$
are both antihermitian, we have
${\a_{\m m}^{\y a}}^*=\a_{\m -m}^{\y a}$ and
${\psi^a_m}^*=\psi^a_{-m}$
under complex conjugation denoted by $*$
in the Majorana representation.

In the action (\ref{D1})
only the traces $\k^{ab}=tr(T^a T^b)$
of quadratic products
of Lie algebra generators
are used,
which are simply
the Killing metric up to normalization.
This is the key property that leads to
the affine Lie group symmetry.

Let $T^a_m=T^a e^{imx/R}$.
They satisfy the loop algebra
\be
[T^a_m, T^b_n]=f^{abc}T^c_{m+n}.
\eq
The trace $\trl$ on the loop algebra
\be
\trl(T^a_m T^b_n)=\k^{ab}\d^0_{m+n}
\eq
is equivalent to
\be
\trl(\cdot)=\int_0^{2\pi R}
\frac{dx}{2\pi R}tr(\cdot).
\eq

By adding the generator $D$ to
the loop algebra according to
\be
[D, T^a_m]=m T^a_m, \label{DT}
\eq
one can rewrite the action (\ref{D1}) in terms of
loop algebra valued quantities
\be
A_{\m}=\a_{\m m}^{\y a}T^a_m,
\quad \Psi=\psi^a_m T^a_m.
\eq
It is
\bee \label{S-loop}
S&=&\int dt\frac{1}{2g}
\trl\left(-\dot{A}_i^2-\dot{A}_1^2
+[\frac{iD}{R}+A_1,A_i]^2
+\frac{1}{2}[A_i,A_j]^2\right. \nn\\
&&\left.+i\Psib(\g^0\dot{\Psi}
+\g^1[\frac{iD}{R}+A_1,\Psi]
+\g^i[A_i,\Psi])\right),
\eqq
where $i=2,3,\cdots,9$.

As a result,
the $U(N)$ gauge symmetry becomes
a loop group symmetry.
This loop group symmetry is no longer
a local symmetry in the usual sense
since there is no coodinate dependence anymore.
It is still a gauge symmetry in the sense that
all observables are required to be
loop group invariants.

It is interesting that
this action can be written as an action
for affine Lie algebra valued quantities.
Define $\hA_i$ and $\hPsi$
to be affine Lie algebra valued quantities.
Let
\be
\hA_i=iz_i K+\a_{im}^{\y a}T^a_m, \quad
\hPsi=i\eta K+\psi^a_m T^a_m,
\eq
where $K$ and $T^a_m$ are affine Lie algebra generators
satisfying
\bee
&[T^a_m, T^b_n]=f^{abc}T^c_{m+n}+m\k^{ab}\d^0_{m+n}K,
\label{aff1}\\
&[T^a_m, K]=0. \label{aff2}
\eqq
The hermitian conjugation on this algebra can be defined by
\be \label{aff3}
K^*=K, \quad {T^a_m}^*=-T^a_{-m}.
\eq
For a more detailed discussion of affine Lie algebras
see for instance Refs.\cite{GO}\cite{Wan}\cite{Fu}.

By adding the generator $D$
one obtains the extended affine Lie algebra,
sometimes simply referred to as
the affine Lie algebra \cite{Wan,Fu}
satisfying (\ref{DT}) and
\be
D^*=D, \quad [D, K]=0.
\eq
The Killing metric
of the affine Lie algebra
is determined algebraically up to normalization
except $Tr(DD)$.
But since the generator $D$ is defined
by the commutation relations and hermiticity
only up to a shift by $K$:
$D\rightarrow D+cK$ for a real number $c$,
we can always make such a shift so that
the result is the following:
\bee
&Tr(T^a_m T^b_n)=\k^{ab}\d^0_{m+n}, \\
&Tr(DK)=Tr(KD)=1, \\
&Tr(DD)=Tr(KK)=0, \\
&Tr(T^a_m D)=Tr(T^a_m K)=0.
\eqq

One can easily check that
the action (\ref{S-loop}) remains the same
if we replace $\trl$ by $Tr$,
$A_{\m}$ and $\Psi$ by $\hA_{\m}$ and $\hPsi$
everywhere in the action.
For configurations with
$\a_{\m m}^{\y a}=\psi^a_m=0$
for all $m\neq 0$ the action
reduces to the effective action of D$0$-branes.

The action of affine Lie algebra valued quantities
is invariant under $K$-shifts:
\be \label{K-shift}
\hA_i\rightarrow \hA_i+a_i K,
\quad \hPsi\rightarrow \hPsi+\lam K,
\eq
hence we can ``gauge'' this symmetry so that
$z_i$ and $\eta$ are not physical observables.
Most importantly, the action is invariant under
the affine Lie group transformation, i.e.,
conjugation of $\hA_i$ and $\hPsi$
by unitary elements in the affine Lie group.
Let $y_1=1$, $y_i=0$ for $i\neq 1$,
the infinitesimal version of this transformation is:
\be \label{affine}
\d_{\eps}\hA_i=[\eps, iy_i D/R+\hA_i], \quad
\d_{\eps}\hPsi=[\eps, \hPsi]
\eq
for a hermitian element in the affine Lie algebra
$\eps= aD+bK+c^a_m T^a_m$.
Explicitly, the coefficients in $\hA_i$ and $\hPsi$
transform as
\bee
&\d_{\eps}z_i=m c^a_m \k^{ab}\a_{i-m}^{\y b},
\label{tr1}\\
&\d_{\eps}\eta=m c^a_m \k^{ab}\th^b_{-m}, \label{tr2}\\
&\d_{\eps}\a_{im}^{\y a}=
ma\a_{im}^{\y a}-im c^a_m y_i/R
+f^{abc}c^b_n\a_{i(m-n)}^{\y c}, \label{tr3}\\
&\d_{\eps}\psi^a_m=ma\psi^a_m
+f^{abc}c^b_n\psi^c_{(m-n)} \label{tr4}.
\eqq
This is equivalent to a loop group transformation
together with a translation in $x$.

A highest weight representation of level $k$
for the affine Lie algebra
is given by a vacuum state satisfying
$T^a_m|0\ra=0$ for $m>0$ and
$K|0\ra=k|0\ra$,
with all other states in this representation
obtained by having products of
generators acting on the vacuum state.
In a highest weight representation,
one can always realize not only the affine Lie algebra
but actually the semi-direct product of
the affine Lie-algebra and the Virasoro algebra,
which is given by (\ref{aff1}-\ref{aff3}) together with
\bee
&[L_m, T^a_n]=-nT^a_{m+n}, \quad [L_m, K]=0, \label{av}\\
&[L_m, L_n]=(m-n)L_{m+n}+\frac{c}{12}m(m^2-1)\d^0_{m+n}.
\eqq
It is called Sugawara's construction,
in which the generators of the Virasoro algebra $L_m$
are relaized by
\be
L_m=\frac{1}{2k+Q}\k_{ab}
(\sum_{n\leq 0}T^a_{m+n}T^b_{-n}
+\sum_{n>0}T^b_{-n}T^a_{m+n}),
\eq
where $Q$ is the quadratic Casimir
in the adjoint representation:
$Q\d^{ab}=f^{acd}f^{bcd}$,
and the central element is
$c=\frac{2kd}{2k+Q}$,
where $d$ is the dimension of the Lie algebra.
>From the commutation relations (\ref{av})
one sees that $L_0$ can be identified with $D$.

When $c=0$, the Virasoro algebra is the algebra
of infinitesimal diffeomorphisms on a circle.
Indeed the transformation of the Virasoro algebra
on $A_{\m}$ and $\Psi$ induced from its action
on the loop algebra (\ref{av}) can be realized by
the diffeomorphism generators
$e^{imx/R}\frac{\del}{\del x}$
acting on the D-strings.
It is tempting to think of
the Virasoro algebra implicit in
highest weight representations
as a signal
of the implicit existence of
conformal field theory (strings)
in the matrix model.

\section{Discussions: Insights into 
Longitudinal Membranes}
\label{discuss}

Previously \IIB/M duality refers to the equivalence
between the M theory compactified on a torus and
type \IIB superstring theory compactified on a circle.
The recently proposed nonperturbative formulation
of M(atrix) theory makes it possible to discuss
the equivalence between the M(atrix) theory 
compactified on a circle and (the strong coupling
limit of) type \IIB theory also compactified on
a circle. In this paper we establish this \IIB/M 
duality in the M(atrix) theory context. Several pieces
of evidence we provide are described in Sec. 
\ref{duality}, and summarized in the abstract 
and introduction.

Here we would like to concentrate on the insights
we have gained into the IMF description of 
M(atrix) theory, since our study involves 
the behavior of a longitudinal membrane.

No one has succeeded in constructing a
longitudinal membrane, in the way a transverse
membrane is constructed \cite{BFSS}, and
no one doubts the existence of a longitudinal
membrane in M(atrix) theory. From our study
we have seen indeed the longitudinal membranes
are hiding in the theory. By the \IIB/M duality
established above, properties of a membrane 
that wraps the longitudinal direction once
can be extracted from those of the D-string 
obtained from compactification on a circle.
These properties are: 
 
\begin{itemize}


\item The excitations of a longitudinal 
membrane are identified with the quantum 
oscillation modes on the D-string, by 
noting that the light-cone energy of 
these modes are finite. This is in sharp
contrast to the energy of a purely
transverse membrane, which comes 
from the commutator potential term.

\item For a longitudinal membrane 
wrapped on the longitudinal torus
with modular parameter $\tau$,
its light-cone energy has a term 
dependent on the ratio $\tau_1/\tau_2$, 
and independent of $P_{11}$, 
as expected from general grounds  
(see eq. (\ref{MM1})). The mechanism
in M(atrix) theory responsible for this
energy is similar to that in a $\theta$ 
vacuum on the D-string, since the latter
contains a topological term with 
$\t_1/\t_2$ as coefficient, analogous 
to the $\theta$ vacuum parameter.

\item The properties of a
longitudinal membrane we extract through 
studying D1-branes are always such that 
they can be viewed as the limit of a 
membrane wrapped on a transverse torus 
with one of the two cycles going to 
infinity. This fact provides us one more 
evidence for eleven dimensional Lorentz 
invariance of M(atrix) theory.

\end{itemize}

In summary, we conclude that the properties
of a membrane which wraps once in the 
longitudinal direction can be extracted
from those of a D-string obtained by 
compactifying M(atrix) theory on a circle. 

Another interesting way to look at the model
considered in this paper is to interchange 
the roles of $R_1$ and $R_{11}$. Namely,
previously we have $R_A=R_1$ 
and $g_A=R_{11}/l_s$; but now we set 
$R'_{A}=R_{11}$ and $g'_{A}=R_1/l_s$.
The ten dimensions of \IIA theory are thus
the $0,2,3,\cdots,9,11$-th dimensions.
If $\t_1=0$, the first two terms in
the spectrum (\ref{MM1}) are the large $P_{11}$
expansion of the relativistic kinetic energy
$\sqrt{P_{11}^2+(mT_0)^2}$ where
$mT_0$ is the mass of $m$ D$0$-branes.
The fourth term in (\ref{MM1}) is
the winding energy of $n$ \IIA string
wound around $R_{11}$ once. With $R_{11}
\to \infty$ and $R_1\to 0$, this model 
should be equivalent to uncompactified 
\IIA theory in the weak 
coupling limit (or its dual \IIB theory).

\section{Acknowledgement}

One of the authors, Y.S.W., thanks Miao Li for
helpful discussions. This work is supported in part by
U.S. NSF grant PHY-9601277.

{\em Note added:} When the writing of this paper is about 
to finish, a preprint of T. Banks and N. Seiberg,
hep-th/9702187, appears which, among other things, 
has some overlapp with part of what we address here.

\appendix

\section{Modular Parameter and Compactification on Torus}

The general idea of compactification of M(atrix) theory
on a compact manifold \cite{BFSS,Tay,GRT}
is to consider the compact manifold
as the quotient of a covering space over a discrete group.
The matrix model on the compact space is then obtained by taking
the quotient of the $U(N)$ matrices over the discrete group,
which is to be embedded in $U(N)$ as a subgroup.

For toroidal compactifications \cite{Tay,GRT},
one needs to choose unitary matrices
$\{U_i, i=1,\cdots,d\}$,
where $d$ is the dimension of the torus,
commuting with each other so that the discrete group
generated by them is isomorphic
to the fundamental group of the torus $\Z^d$.
The action of $U_i$ on the coordinates is
\be
U_i X^{\m} U_i^{\dagger}=X^{\m}+e_i^{\m},
\eq
where $e_i^{\m}$ is the basis of the lattice
whose unit cell is the torus.
Obviously $U_i$ is the operator
translating all fields a whole cycle along $e_i$.

The matrix theory is then restricted to
be invariant under the action of the $U_i$'s.
This is most easily realized by viewing
$e^i_{\m}X^{\m}$ as the covariant derivative
$\frac{\del}{\del x^i}+A_i$ in the direction of $e^i$,
dual to $e_i$,
and the $U_i$ as $e^{2\pi i x^i}$,
where $x^i$ is a coordinate on the dual torus
and so we have interchanged the role
of coordinates and momenta.
Naturally the matrix model becomes
a $d+1$ dimensional gauge field theory.

For a slanted two-torus with modular parameter
$\t=\t_1+i\t_2$ and radii $R_1, R_2=\t_2 R_1$,
it is convenient to introduce slanted coordinates
$(X'_1, X'_2)$ by
\bee\label{XX'}
\left(\begin{array}{c}
X'_1 \\
X'_2 \end{array}\right)=
\left(\begin{array}{cc}
1 & -\frac{\t_1}{\t_2} \\
0 & 1 \end{array}\right)
\left(\begin{array}{c}
X_1 \\
X_2 \end{array}\right)
\eqq
out of the orthonormal coordinate system $(X_1, X_2)$.
The nice thing about the $X'_i$'s is that
they live on circles of radii $R_i$.
The discrete group generated by $U_1, U_2$
with $U_1 U_2=U_2 U_1$ acts on them simply as
$U_i X'_j U_i^{\dagger}=X'_j+2\pi\d_{ij}R_j$.
>From the action $\frac{1}{2}(\dot{X_1}^2+\dot{X_2}^2)$
one defines conjugate variables and finds
\be
\left(\begin{array}{cc}
P'_1 & P'_2 \end{array}\right)=
\left(\begin{array}{cc}
P_1 & P_2 \end{array}\right)
\left(\begin{array}{cc}
1 & \frac{\t_1}{\t_2} \\
0 & 1 \end{array}\right).
\eq

It follows that the kinetic energy is
$\frac{1}{2}(P_1^2+P_2^2)=
\frac{1}{2}\left({P'_1}^2+(P'_2-\frac{\t_1}{\t_2}P'_1)^2\right)$,
giving the spectrum of
$\left(\frac{m_1}{R_1}\right)^2
+\left(\frac{m_2-\t_1 m_1}{R_2}\right)^2$.

The case of a longitudial membrane can be 
inferred from this result. To do so, note that 
$R_1$ and $R_2$ here correspond to $R_{11}$ and 
$R_1$, respectively,
in this paper for the longitudinal torus.
Therefore the first two terms in eq. (\ref{MM}) follow.
Further, the action $\frac{1}{2}(\dot{X_1}^2+\dot{X_2}^2)$
can be written as
$\frac{1}{2}(\dot{Y}_1^2+(\dot{Y}_2+\t_1\dot{Y}_1/\t_2)^2)$,
where $Y_1=\t_2 X'_1/|\t|$ and $Y_2=|\t| X'_2/\t_2$.
Applying this result to the longitudinal case
in the infinite momentum frame ($\dot{Y_1}=1$),
we see that the prescription is to
replace $R_{11}$ by $R'_{11}$
and $\dot{A_1}$ by $(\dot{A'_1}+i(\t_1/\t_2)T_s)$
where $A'_1$ (corresponding to $iT_s Y_2$)
is valued in the range $iT_s[0, 2\pi R'_1)$.
This gives what we found in Sec.\ref{spectrum}.

\baselineskip 22pt

\bibliography{IIBM}
\bibliographystyle{unsrt}

\end{document}